\documentclass[10pt, conference, letterpaper]{IEEEtran}

\usepackage{graphicx, xcolor, colortbl}
\usepackage{algorithm, algpseudocode, subfigure}
\usepackage{url,amsmath,amssymb,array,threeparttable,multirow,booktabs}
\usepackage[sort&compress,square,comma,numbers]{natbib}
\usepackage{multicol, multirow}

\newcolumntype{P}[1]{>{\centering\arraybackslash}p{#1}}

\date{October 2023}

\begin{document}
\title{Power-Efficient Indoor Localization Using Adaptive Channel-aware Ultra-wideband DL-TDOA}
\author{
\IEEEauthorblockN{
Sagnik Bhattacharya, Junyoung Choi, Joohyun Lee}
\IEEEauthorblockA{
\textit{Communications Standards Research Team, Samsung Electronics, Seoul, South Korea} \\
Email: \{sagnik.b, juny.choi, jh5648.lee\}@samsung.com}
}

\maketitle

\begin{abstract}
Among the various Ultra-wideband (UWB) ranging methods, the absence of uplink communication or centralized computation makes downlink time-difference-of-arrival (DL-TDOA) localization the most suitable for large-scale industrial deployments. However, temporary or permanent obstacles in the deployment region often lead to non-line-of-sight (NLOS) channel path and signal outage effects, which result in localization errors. Prior research has addressed this problem by increasing the ranging frequency, which leads to a heavy increase in the user device power consumption. It also does not contribute to any increase in localization accuracy under line-of-sight (LOS) conditions. In this paper, we propose and implement a novel low-power channel-aware dynamic frequency DL-TDOA ranging algorithm. 
It comprises NLOS probability predictor based on a convolutional neural network~(CNN), a dynamic ranging frequency control module, and an IMU sensor-based ranging filter. 
Based on the conducted experiments, we show that the proposed algorithm achieves 50\% higher accuracy in NLOS conditions while having 46\% lower power consumption in LOS conditions compared to baseline methods from prior research. 
\end{abstract}


\vspace{-5pt}

\section{Introduction}
\label{sec:intro}
Ultra-wideband (UWB) spectrum-based ranging offers high accuracy (e.g., 15~cm) and it makes a topic of high research interest in recent years. 
The most prominently used UWB-based ranging methods are double-sided two way ranging (DS-TWR) and downlink-time-difference-of-arrival (DL-TDOA)~\cite{2023-wcnc-sagnik}. 
In DL-TDOA methods, there are pre-assigned clusters of UWB anchors. 
The anchors in a cluster regularly exchange UWB messages containing timestamps and reply time information. 
A user device in that area listens to those messages, extracts the timestamp information, and triangulates its location coordinates in a downlink manner, without requiring any uplink communication or centralized computation. 
This makes the DL-TDOA based localization highly scalable with increasing number of user devices, and thus the perfect candidate for large-scale industrial deployments. 
However, the presence of temporary obstacles (e.g. moving humans/cars) or permanent obstacles (e.g. walls/pillars in the deployment scenario) often lead to signal blockage effects, hence resulting in non-line-of-sight (NLOS) channel paths. 
This is detrimental to the localization performance, and causes time synchronization errors and localization errors. 
This problem is exacerbated in highly mobile or crowded situations. 

Prior research handles the problem by increasing the DL-TDOA ranging frequency, i.e. the number of times DL-TDOA-based localization is conducted in a given time. 
However, that leads to a heavy increase in user device power consumption, provides very small performance improvement in NLOS conditions, and no performance gain in LOS conditions. 
Another research deals with the problem by introducing an auxiliary ranging process based on time-of-flight (TOF) along with the primary DL-TDOA process~\cite{tdoa&tof}.
While this improves the localization accuracy to a higher degree, especially in NLOS conditions, the process is not scalable to many user devices because the DS-TWR process involves one-to-one communication between each anchor and each user device.

The ability of deep learning-based architectures to extract meaningful patterns from given data, and use it to achieve near-optimal efficient solutions to otherwise intractable problems in real-time, has led to its widespread recent use in wireless optimization problems~\cite{2022-spcom-sagnik, kalman-deep-learning}. Convolutional neural networks (CNN) are extremely efficient in extracting practical spatial or temporal features directly from raw data~\cite{2022-spcom-sagnik}, and are thus extremely useful in our wireless scenario.

In this paper, we propose a novel power-efficient channel-aware adaptive UWB DL-TDOA ranging algorithm. 
The proposed algorithm comprises NLOS probability predictor using a convolutional neural network (CNN) for the received UWB signals, a dynamic DL-TDOA frequency control module, and an IMU sensor-based ranging filter. 
We also adopt the localization predictor based on the recurrent neural network (RNN) from \cite{2023-iccws-sagnik} to further lower the required DL-TDOA ranging frequency. 
The proposed algorithm is highly power-efficient, as well as scalable, since it does not involve DS-TWR. 
We conduct localization experiments on Samsung Galaxy S21 Ultra (GS21U) smartphones, connected to Qorvo Decawave UWB chipset. 
We show that the proposed algorithm achieves 50.4\% higher localization accuracy, compared to baseline methods, in predominantly NLOS conditions, while having 46.3\% lower power consumption in LOS conditions.

Specifically, our novel contributions in this paper are the following:

\begin{itemize}
    \item We propose a novel CNN-based algorithm which takes as its input the received signal channel impulse response (CIR) of the various anchors which are currently participating in the DL-TDOA localization process, and predicts the NLOS probability for each of those anchors.
    \item We design a novel dynamic DL-TDOA ranging frequency module, which adjusts the user device ranging frequency according to the average channel quality. In addition, we also design an IMU sensor-based filter, which turns off the DL-TDOA ranging module when the user is not moving. These two modules help reduce power consumption.
    \item To keep the localization accuracy unaltered during the times when the dynamic ranging frequency is reduced, we use a recurrent neural network (RNN)-based future localization predictor.
    \item To verify the practical applicability of our proposed module architecture, we deploy it on Samsung GS21U smartphones connected to Qorvo UWB chipsets. Compared to prior work, we achieve $50.4\%$ higher localization accuracy in NLOS conditions, while having $46.3\%$ lower power consumption in LOS conditions. 
\end{itemize}

\section{Related Work}
\label{sec:related}
There has been significant prior research on improving the performance of UWB DL-TDOA-based localization~\cite{kalman1, kalman2, tdoa&tof, kalman-uwb-imu-survey}. However, those works are primarily concerned with either improving the localization accuracy~\cite{kalman1, kalman2}, or reducing the power consumption~\cite{tdoa&tof}. The authors in~\cite{kalman1} use a Kalman Filter-based approach to improve the accuracy of time-of-arrival (TOA)-based localization. Even though this is a low power consumption algorithm, the Kalman Filter-predicted localization coordinates keep diverging from the ground truth values in predominantly NLOS situations. This is because of the Kalman Filter's gaussian noise assumption, which is not true for the IMU sensor (accelerometer and gyroscope) data, where the error has a bias and accumulates over time. The authors in \cite{tdoa&tof} use a combination of TDOA and time-of-flight (TOF)-based measurements during localization. They achieve a $75\%$ user device-side power consumption reduction compared to TOF-based localization, while maintaining the same localization accuracy as baseline methods. The usage of TOF requires the anchors to respond to individual user devices using response messages. Hence, this algorithm is not scalable to many user devices. The authors in~\cite{kalman-deep-learning} use deep learning on the UWB channel impulse response (CIR) to estimate the probabilities of various factors which are representative of LOS/NLOS conditions. While this method helps improve localization accuracy, it also involves anchor-to-user device communication, which affects its scalability. 
\section{Preliminaries of DL-TDOA localization}
\label{sec:preliminary}

\begin{figure}
    \centering
    \includegraphics[width=0.7\columnwidth]{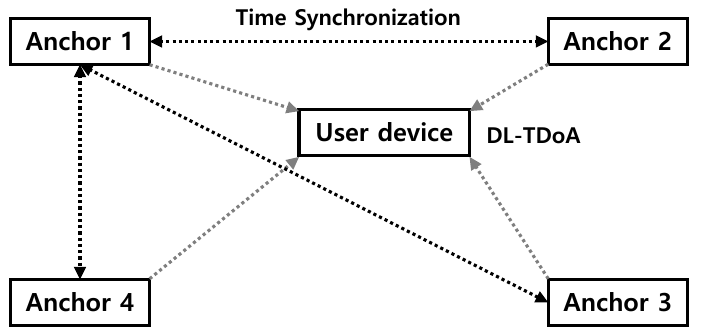}
    \caption{A DL-TDOA cluster comprising 4 anchors and a user device}
    \label{fig:DL-TDOA}
\end{figure}

Fig.~\ref{fig:DL-TDOA} shows a DL-TDOA ranging cluster. The UWB anchors are fixed at pre-defined locations and are assigned clusters~\cite{2023-wcnc-sagnik}. We consider $M$ anchors, which are all assigned the same cluster. These $M$ anchors periodically exchange UWB messages containing timestamps. These messages are used by the anchors to maintain time synchronization, which is crucial for successful DL-TDOA localization. The user device, equipped with a UWB-enabled chipset, ``listens" to these messages and collects these timestamps. The user device then calculates the TDOA values, and triangulates its location using least-squares estimation. This eliminates the need of any uplink communication or computation at the anchor. The presence of NLOS/multipath effects between the user device and the anchor leads to inaccuracies in the measurements of reply time, and hence the TDOA values. This adversely affects the least squares-based estimation and leads to high localization errors.

\section{Proposed Algorithm}
\label{sec:proposed}

\begin{figure}[t]
    \centering
    \includegraphics[width=\columnwidth]{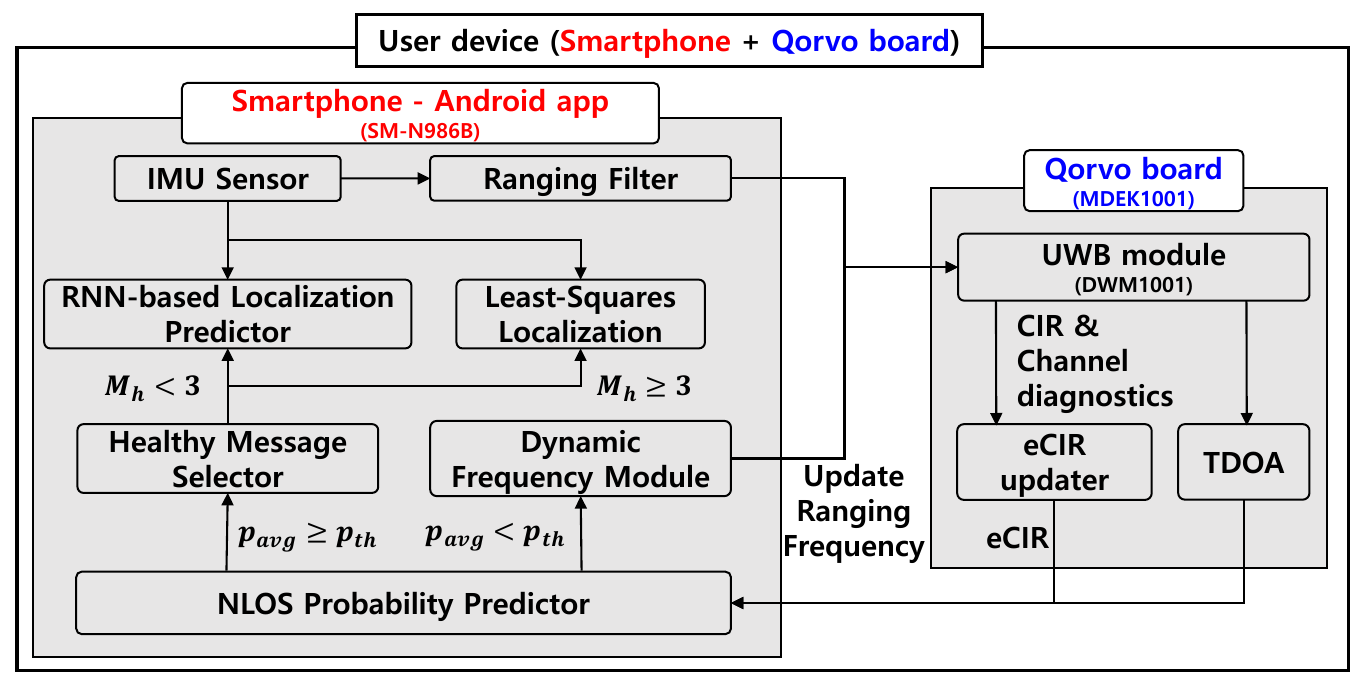}
    \caption{The algorithmic flow for adaptive DL-TDOA localization.}
    \label{fig:proposed-model}
    \vspace{-10pt}
\end{figure}

\begin{figure}[t]
    \centering
    \subfigure[The CIR divided into signal and noise based on maxNoise]{
    \includegraphics[width=0.444\columnwidth]{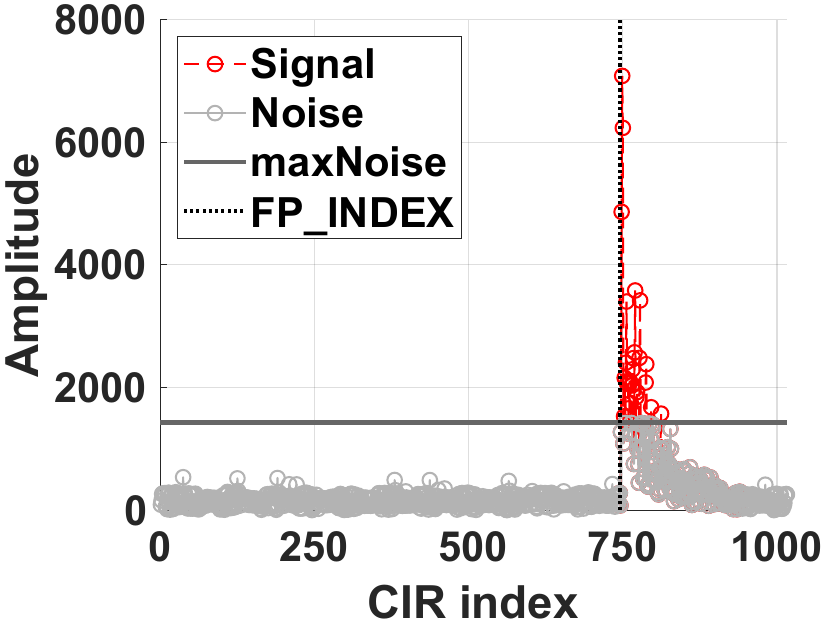}
    \label{fig:cir1016}
    }
    \subfigure[The difference in the amplitude and the number of peaks]{
    \includegraphics[width=0.444\columnwidth]{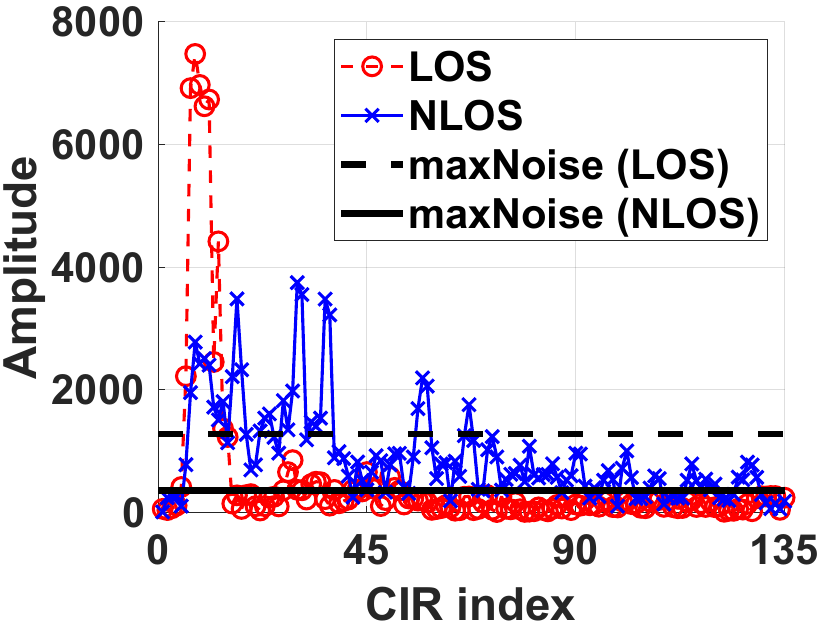}
    \label{fig:cir200}
    }
    \vspace{-2mm}
    \caption{The example of effective CIR and difference between LOS/NLOS.}
    \label{fig:eCIR}
\end{figure}

\begin{figure}[t]
    \centering
    \subfigure[LOS/NLOS classification model architecture with 4 CNN layers]{
    \includegraphics[width=0.9\columnwidth]{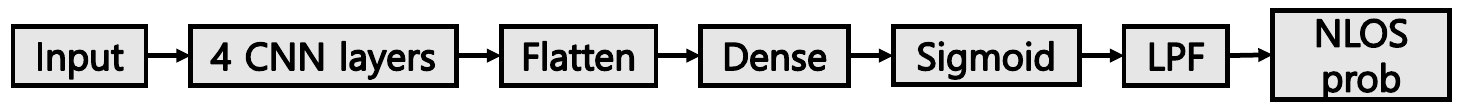}
    \label{fig:nlos_classifier_architecture}
    }
    \subfigure[CNN layer architecture]{
    \includegraphics[width=0.9\columnwidth]{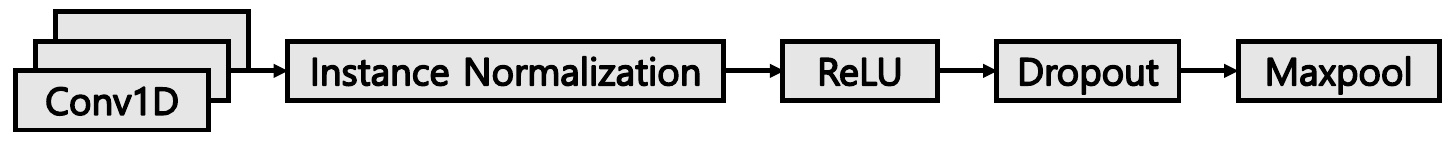}
    \label{fig:nlos_classifier_cnn}
    }
    \vspace{-2mm}
    \caption{CNN based model for LOS/NLOS classification using CIR input.}
    \label{fig:nlos_classifier}
\end{figure}

\begin{figure}
    \centering
    \includegraphics[width=1\columnwidth]{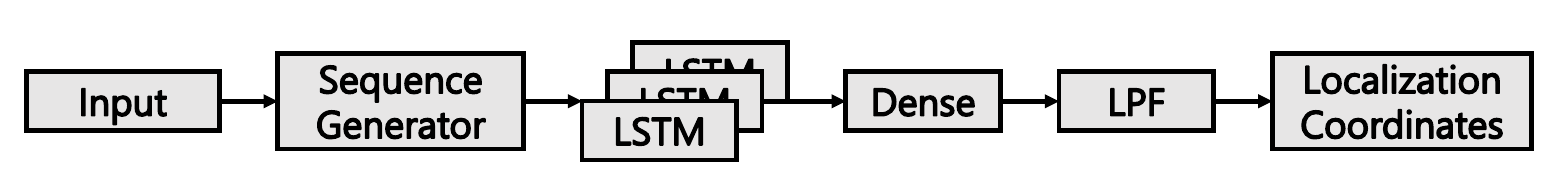}
    \caption{The RNN model based LSTM for estimating the location.}
    \label{fig:rnn}
\end{figure}

Fig.~\ref{fig:proposed-model} shows the proposed model for power-efficient channel-aware DL-TDOA-based localization. We assume that there are $M$ anchors inside a cluster involved in the DL-TDOA process. The UWB module in the user device collects the UWB CIR and timestamps from $M$ anchors at the specified frequency. The first step comprises a CNN-based NLOS probability predictor. It operates on the received CIR from each anchor and predicts the NLOS probability for the channel path between that anchor and the user device. If the average of those NLOS probabilities, namely $p_{avg}$, is less than a pre-specified threshold $p_{th}$ (=0.5) the situation is classified as predominantly LOS, and predominantly NLOS otherwise. 

In the first case, i.e., the case of a predominantly LOS situation, the outputs from the CNN are sent to the dynamic frequency module, which updates the frequency at which the user device conducts DL-TDOA-based localization. This resulting updated frequency is sent to the UWB module. The IMU sensor-based ranging filter also sends its output to the UWB module. These two inputs together determine the next time instant when the UWB data is collected. Finally, the DL-TDOA data collected from the UWB module is sent to the localization module, which calculates the localization coordinates of the user device using the least squares estimation~\cite{least-squares}. 

In the second case, i.e., the predominantly NLOS situation, the NLOS probabilities are sent to the healthy message selection module, which selects a subset of the UWB messages collected from the $M$ anchors, which correspond to LOS channel paths. If the total number of messages (each message from a different anchor) in that subset is less than $3$, then the entire ranging data for that round is discarded. If that is not the case, then the selected DL-TDOA data in that subset are sent to the localization module. In addition to the localization module, we adopt the RNN-based robust localization predictor from~\cite{2023-iccws-sagnik}. This helps provide uninterrupted localization data in cases where (1) the original collected UWB data is discarded as mentioned above, or (2) the frequency of the UWB data collection is reduced by the dynamic frequency module in a predominantly LOS condition. 

The individual modules in the algorithmic flow are explained in detail as follows:

\subsection{Effective CIR (eCIR) Updater}
\label{subsec:eCIR}
The eCIR collection module is the first step in the proposed model, which receives the received raw CIR data in the user device from the $M$ UWB anchors. Fig.~\ref{fig:cir1016} shows a typical received CIR, which contains 1016 amplitude values as a function of time. Here, FP\_INDEX represents the first path index or the beginning of the first path of the received CIR. All values prior to FP\_INDEX are irrelevant and hence removed. Here, maxNoise represents the noise floor, hence all received values below that are essentially noise. Hence, we see that all the essential and relevant CIR information is confined to $200$ values after the FP\_INDEX, and hence we select only those $200$ values as the effective CIR or the eCIR. The eCIR is sent to the NLOS probability predictor.

\subsection{NLOS Probability Predictor}
\label{subsec:nlos-prob}
The NLOS probability predictor module, which is the first step of the proposed algorithm, ingests the received eCIRs from the eCIR module. Each of the received eCIRs are sent individually to the CNN-based model, which predicts the corresponding NLOS signal probability. Fig.~\ref{fig:nlos_classifier_architecture} shows the neural network model architecture used for the NLOS probability prediction. The input layer ingests the individual eCIRs. It is followed by 4 CNN layers, which are efficient at meaningful feature extraction. The output from the last CNN layer is sent to the flatten layer, which converts it into a 1-dimensional array. This is taken as input by the dense layer, in which all neurons of the current layer are connected with all neurons of the previous layer using weights. The output of the dense layer is passed into the sigmoid activation layer, which applies the sigmoid function and produces a number between 0 and 1. This number is then passed through a low pass filter (LPF), which applies a weighted mean of the current input and the previous stored output. Finally, the output of the LPF represents the probability of the given input CIR corresponding to a NLOS signal.

Fig.~\ref{fig:nlos_classifier_cnn} shows the neural network layers constituting an individual CNN layer. The four Conv1D layers each comprise a pre-defined number of 1 dimensional vectors, named filters. All of the filters have a pre-defined size called kernel size. These filters help extract meaningful spatial/temporal features from the input data. Each of these 1 dimensional filters is convolved with the input to produce a 1 dimensional vector. These vectors are stacked together to form a 2 dimensional matrix, which is then passed to the instance normalization (IN) layer. The IN layer normalizes the incoming input along a pre-specified axis. This, unlike the more traditional batch normalization, in insensitive to training data statistics and hence helps prevent overfitting. The rectified linear unit (ReLU) layer ingests the normalized output of the IN layer, and applies the ReLU function on it elementwise. Next, the dropout layer randomly masks the input from a pre-defined percentage of neurons at runtime. This helps reduce overfitting caused by training data reliance. The final layer, which is the maxpool layer, also has a pre-defined kernel size. It applies the maxpool function, i.e., replacing a contiguous set of elements with the maximum among them, along a specified axis. This helps reduce the dimensionality of the data, which boosts efficiency and helps reduce overfitting.

\subsection{Dynamic Ranging Frequency Module}
\label{subsec:dynamic-frequency}
In the predominantly LOS situation, the dynamic frequency module updates the DL-TDOA ranging frequency $f_{dyn}$. We calculate an intermediate value $f^{'}$ which is given by $f^{'}=\frac{f}{\lfloor \frac{p_{th}}{p_{avg}} \rfloor}$, where $f$ is the default ranging frequency of $5$Hz. $f_{dyn}$ is given as
\begin{align}
    f_{dyn} = 
    \begin{cases} 
    0.1 Hz & \text{if } f^{'} < 0.1 \\
    5 Hz & \text{if } f^{'} > 5 \\
    f^{'} & \text{otherwise }
    \end{cases}
    \label{eq:dynamic-frequency}
\end{align}
In predominantly LOS situations, i.e., when the average NLOS probability $p_{avg} < p_{th}$, the dynamic frequency module reduces the ranging frequency as per Eq.~\eqref{eq:dynamic-frequency}. The minimum and maximum allowed frequencies are $0.1$~Hz and $5$~Hz, respectively. This leads to a reduction of power consumption.

\subsection{Healthy Message Selector}
\label{subsec:healthy}
In a predominantly NLOS situation, the received TDOA values, along with their individual NLOS probabilities, are sent to the Healthy Message Selector module. This module selects a subset of the TDOA values, for which the corresponding average NLOS probability $p^{'}_{avg} \leq p_{th}$. To do this, it iteratively eliminates the TDOA from the set, for which the NLOS probability is the highest, and then updates the new average NLOS probability of the remaining TDOA values. If the number of TDOA values in that final subset is $3$ or more, which is the minimum number required for 2D localization, then the situation is determined to be ``partially LOS". In the other case, i.e., if the number of remaining TDOA values in the final set is $2$ or less, the situation is determined to be ``fully NLOS". In a partially LOS situation, the subset of TDOA values is sent to the Least Squares Estimation-based Localization module. In a fully NLOS situation, all the TDOA values are discarded, and the RNN-based Localization Predictor module is used.

\subsection{RNN-based Localization Predictor}
\label{subsec:rnn}
Fig.~\ref{fig:rnn} shows the RNN model architecture used for future localization coordinates prediction. The input comprises the UWB TDOA values and the IMU sensor (accelerometer and gyroscope) data as a continuous incoming stream. The sequence generator stacks the UWB TDOA and IMU values for the past $N_{in}$ timesteps and generates sequences of dimension $(N_{in}, (M-1)+3+3)$. Here, $M$ is the number of anchors currently participating in the DL-TDOA ranging session. Hence, we have $M-1$ TDOA values. The IMU sensor, i.e. the accelerometer and gyroscope readings along the X, Y, and Z axes contribute another $6(=3+3)$ values. The generated sequence is sent as input to the RNN. The RNN comprises a long short-term memory (LSTM) encoder-decoder structure. LSTM cells are extremely efficient in extracting practical temporal information from given time series data. The encoder architecture comprises a set of LSTM blocks, which ingest the input sequence and generate compressed embeddings. The decoder again consists of another set of LSTM blocks which ingest the generated embeddings and provide an output of dimensions $(N_{out}, 2)$. This is followed by a dense layer. Finally, the output from the dense layer is passed through a low pass filter (LPF) to smoothen out random fluctuations in localization coordinates. The output of the LPF, which is of dimension $(N_{out}, 2)$ provides the predicted localization coordinates for the next $N_{out}$ timesteps. 

It is to be noted that, as a result of the RNN-based localization prediction, we always have the predicted localization coordinates for $N_{out}$ future timesteps available to us. During the DL-TDOA localization process, there are cases of signal reception failure, which lead to non-availability of collected TDOA values. In such cases, the predicted localization coordinates at that timestep are used to calculate the predicted TDOA values. For example, suppose the pre-provided fixed locations of the $M$ anchors in the current cluster are given by $\{(x_i, y_{i}) ~|~ i \in \mathbb{Z}, 1\leq i\leq M\}$, where the first anchor ($i=1$) is the initiator. If the RNN-predicted localization coordinates of the user device for a particular future timestep are $(x_{ud}, y_{ud})$, then the predicted TDOA value corresponding to the $j^{th}$ responder, $\alpha_{j}$ is given by:
\begin{multline}
     \alpha_{j} = \frac{1}{c} (|\sqrt{(x_j-x_{ud})^2+(y_j-y_{ud})^2} - \\
     \sqrt{(x_1-x_{ud})^2+(y_1-y_{ud})^2}|) ~\forall j \in \mathbb{Z}, 2 \leq j \leq M
\end{multline}
where $c$ is the speed of light in vacuum.
These predicted (or ``augmented") TDOA values can be used as input to the RNN at the corresponding future timestep, if the Qorvo chipset fails to obtain external UWB signal at that time. Hence, we always obtain localization predictions using the RNN, even during signal reception failures. This perennial supply of augmented TDOA values, along with the IMU sensor data, makes the localizer robust to multipath effects and penetration losses.

\section{Baseline Methods}
\label{sec:baseline}
We compare the performance of our proposed power-efficient high-accuracy UWB DL-TDOA localization algorithm to several baseline methods from prior research. Baseline 1 includes the traditional least squares estimation-based DL-TDOA localization~\cite{least-squares}. This method has a static ranging frequency of $200~ms$. It also lacks any predictive power, i.e., it fails to compute localization coordinates in the absence of UWB signals from the minimum number of anchors that is necessary for triangulation, i.e. $3$. Our baseline 2 is from~\cite{tdoa&tof}, which uses a combination of UWB TOF and DL-TDOA values for localization. Baseline 3 is adopted from~\cite{kalman-deep-learning}. The algorithm used in this paper infers several LOS/NLOS-related channel conditions using deep learning on the received CIR in two-way-ranging (TWR), and then accordingly adjusts the Kalman filter for localization prediction and estimation.
\section{Experimental Results}
\label{sec:performance}

\begin{figure}
    \centering
    \includegraphics[height=5cm]{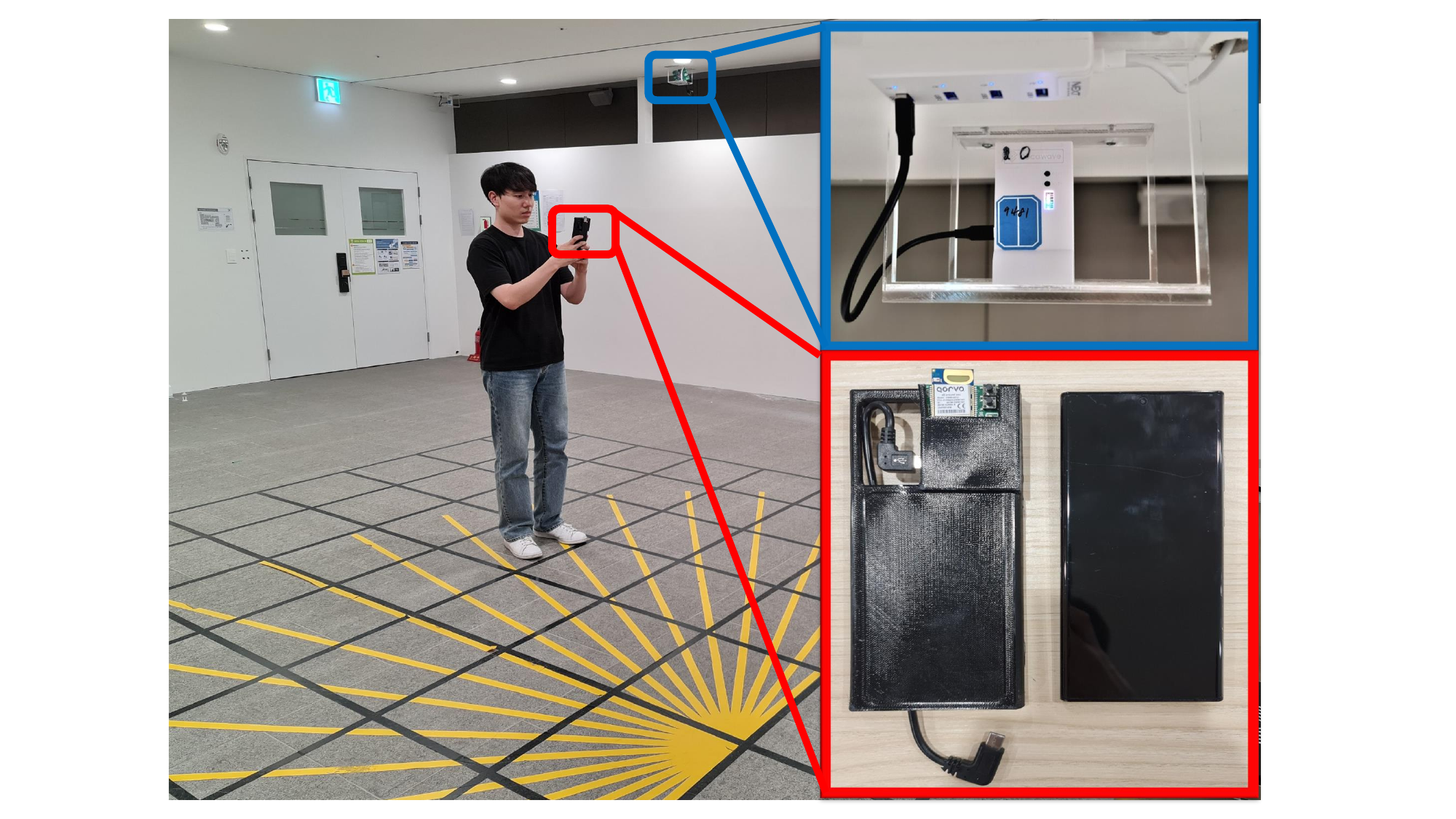}
    \caption{The real-world experiment setting, showing the user holding the device, and the anchor installed on the ceiling.}
    \label{fig:setup}
\end{figure}

\begin{table}[t]
    \caption{UWB and IMU sensor configuration parameters}
    \centering
    \begin{tabular}{|c|c|}
    \hline
       \textbf{Parameter}  &  \textbf{Value}\\ \hline
        UWB Ranging Block Duration & $200$ms \\ \hline
        UWB Ranging & DL-TDOA \\ \hline
        UWB Mode & Multicast \\ \hline
        No. of Anchors & $7$ \\ \hline
        IMU Sensor Frequency & $16$Hz \\ \hline
    \end{tabular}
    \label{tab:ranging-imu-params}
    \vspace{-3mm}
\end{table}

\begin{table}[t]
    \caption{Model Parameters}
    \centering
    \begin{tabular}{|c|c|}
    \hline
         \textbf{CNN hyperparameter} & \textbf{Value} \\ \hline
         Conv layer \{1,2,3,4\} kernel size & \{9,13,17,17\} \\ \hline
         Conv layer \{1,2,3,4\} no. of filters & \{64,64,128,256\} \\ \hline
         Maxpool kernel size & 2 \\ \hline
         Dropout rate & 0.1 \\ \hline
         Dense layer no. of neurons & 64 \\ \hline
         Optimizer & Adam \\ \hline
         Loss function & Binary cross-entropy \\ \hline
         Batch size & 50 \\ \hline
         Maximum no. of training epochs & 100 \\ \hline
         \textbf {RNN hyperparameter} & \textbf {Value} \\ \hline
         $N_{in}$ & 16 \\ \hline
         LSTM layer \{1,2,3,4\} no. of units & \{32,32,32,64\} \\ \hline
         Dense layer no. of neurons & 64 \\ \hline
         Optimizer & Adam \\ \hline
         Loss function & Mean-squared error \\ \hline
         Batch size & 10 \\ \hline
         Maximum no. of training epochs & 50 \\ \hline
    \end{tabular}
    \label{tab:model-parameters}
\end{table}

\subsection{Simulation Setup}
\label{subsec:setup}
Fig.~\ref{fig:setup} shows the simulation setup for the performance evaluation of the proposed algorithm. The lab environment is rectangular of dimensions $9~m \times 6~m$, and contains $7$ Qorvo anchors placed on the ceiling at a height of $2.7~m$ as shown in Fig.~\ref{fig:setup}. The user device is Samsung GS21, which is connected to a Qorvo chipset using a USB port. 
Squares marked on the floor surface with pre-provided location coordinates provide the ground truth localization data. The user carries the Samsung S20U device, attached to the Qorvo chipset by cable, in a handheld manner. The Android app running on the device collects the TDOA, CIR, and IMU sensor data. 

\subsection{Model Implementation}
\label{subsec:implement}
\noindent\textbf{User device:}
We implement the application on the commercial smartphone (i.e., SM-N986B) running on Android~11 to show the feasibility.
Both the CNN-based NLOS probability predictor and RNN-based localization predictor are implemented using TensorFlow 2.10.0 and converted to TensorFlow-Lite for mobile deployment. Other modules including IMU are implemented using Android Studio.

\noindent\textbf{UWB:}
We implement the UWB module on the user device based on the example code from the Qorvo DWM1001~\cite{dwm1001example}. 
The DL-TDOA operation in Qorvo board is implemented by following the user manual using SEGGER Embedded Studio for ARM 5.32a~\cite{dw1000user, seggerARM}. The UWB ranging and IMU sensor parameters used for configuring the user device and the installed anchors are shown in Table~\ref{tab:ranging-imu-params}.

\subsection{Model Training}
\label{subsec:train}
For training the CNN-based NLOS probability predictor model, we collect volunteer data in both LOS and NLOS conditions. This CIR data for training is collected at a default frequency of $5$~Hz, and we collect a total of $4000$ LOS samples and $4000$ NLOS samples. To prevent the overfitting scenario associated with training data shortage, this collected dataset is combined with publicly available open datasets.

For the RNN-based localization predictor, we collect training data and train the model as in~\cite{2023-iccws-sagnik}. The UWB DL-TDOA data (along with the CIR data) is collected by the Qorvo module at a frequency of $5$Hz. We can choose the value of $N_{in}$, which is the number of timesteps for which we send the TDOA data to the RNN. We experiment with various values of $N_{in}$ and obtain the best localization performance for $N_{in}=16$. The IMU sensor data is also sent as input to the RNN. We collect new IMU sensor readings every $62.5$~ms. Assuming that the average human walking speed is 2 steps/second, and that we want to send the IMU data corresponding to the latest 2 steps into the RNN, we send 1 second's worth of IMU data. We want to keep the timestep dimension of the UWB DL-TDOA and IMU data to be the same, i.e. $16$. Hence we should send $16$ IMU data samples, covering the data for $1~s$. Hence, the required IMU data interval is $1/16=62.5~ms$. The various parameters used for the proposed model implementation are showed in Table~\ref{tab:model-parameters}.

\begin{figure}[t]
    \centering
    \includegraphics[trim = 100 240 100 250 , clip, width=0.81\columnwidth]{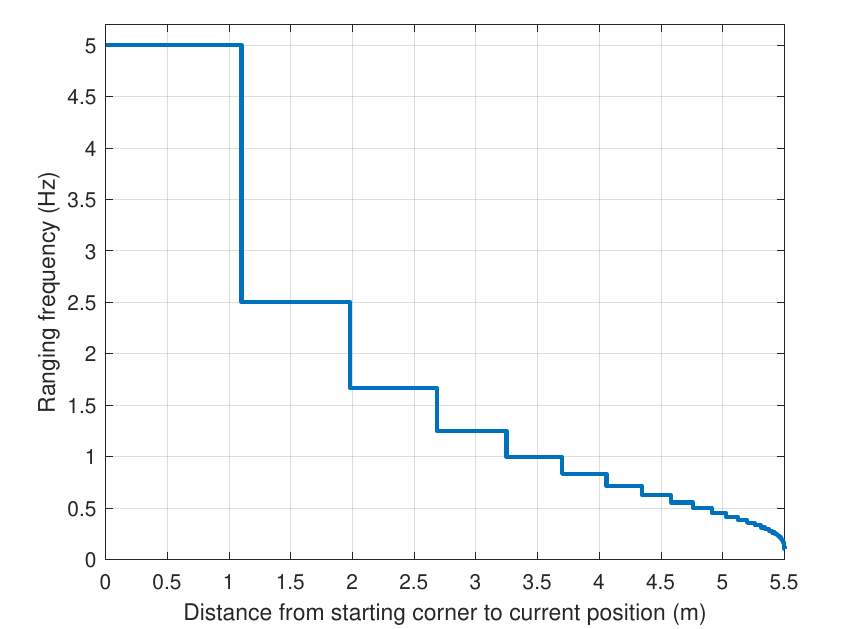}
    \caption{Change in UWB DL-TDOA frequency, as dictated by the dynamic ranging frequency module, with change in average NLOS probability.}
    \label{fig:dynamic-frequency}
    \vspace{-3mm}
\end{figure}

\subsection{Performance Evaluation}
\label{subsec:results}
Fig.~\ref{fig:dynamic-frequency} shows the change in DL-TDOA ranging frequency with the change in the NLOS probability as predicted by the algorithm. The X-axis values increase from $0$, indicating one corner of the rectangular lab, i.e., $(9~m,6~m)$ to $\sqrt{(9/2)^2+(6/2)^2}=5.41~m$, indicating the center of the rectangle. As the user device moves along a straight line from one corner of the cluster toward the center, the scenario changes from a predominantly NLOS situation to a predominantly LOS situation. As a result, the dynamic ranging frequency keeps decreasing, leading to lower power consumption. The piece-wise nature of the graph is because of the floor function in the definition of $f^{'}$ in Eq.~\eqref{eq:dynamic-frequency}.

\begin{table}[t]
    \caption{Average power consumption and average localization error achieved by proposed and baseline algorithms}
    \centering
    \begin{tabular}{|P{2.5cm}|P{1.9cm}|P{1.8cm}|}
    \hline 
    & \textbf{Average Current (mA)} & \textbf{Localization Error (cm)} \\  \hline
    \textbf{LOS Scenario} & & \\ \hline
    Least Squares  & 30.8 & 3.5 \\ \hline 
    TDOA+TOF & 45.7 & 2.8 \\ \hline
    NLOS Prob+KF & 36.4 & 3.3\\ \hline
    Proposed Algorithm  & 19.7 & 3.3 \\ \hline 
    \textbf{NLOS Scenario} & & \\ \hline
    Least Squares  & 30.8 & 17.6 \\ \hline 
    TDOA+TOF & 45.7 & 12.8\\ \hline
    NLOS Prob+KF & 41.3 & 11.5 \\ \hline
    Proposed Algorithm  & 28.2 & 6.7 \\ \hline 
    \end{tabular}
    \label{tab:results}
    \vspace{-5mm}
\end{table}

Table~\ref{tab:results} shows the average localization error and the average user device current consumption per UWB ranging block, obtained during UWB DL-TDOA localization, using the proposed and baseline algorithms. As explained in Section~\ref{sec:baseline}, baselines 1 (Least Squares), 2 (TDOA+TOF) and 3 (NLOS Prob+KF) refer to the least-squares estimation, the combination of TDOA and TOF methods, and the deep learning-based adaptive kalman filter methods respectively. Table~\ref{tab:results} is divided into two parts, corresponding to the experiment scenario being predominantly LOS or predominantly NLOS. The predominantly LOS situation is the ideal lab environment in which the user walks around with the smartphone in a handheld manner. The predominantly NLOS situation is created by partially blocking all of the anchors in the lab. From Table~\ref{tab:results}, we see that in predominantly LOS conditions, the localization error achieved by all the algorithms are quite similar, with Baseline 2 being slightly better. This is because it uses a combination of TDOA and TOF values, leveraging the higher accuracy of TOF-based localization. As for the average current consumption, Baseline 2 performs worst. This is because of the message-heavy two-way-ranging involved in the TOF phase, where there is one-to-one communication between the user device and each anchor, unlike the DL-TDOA process. Baseline 3 uses deep learning-based NLOS classification and accordingly updates the Kalman Filter (KF) parameters for localization prediction. Although it uses DL-TDOA localization, it is computationally heavy, and thus has a higher current consumption than the traditional least-squares estimation-based DL-TDOA localization, i.e. Baseline 1. Finally, the proposed algorithm has the lowest current consumption of $19.7$~mA. This is attributed to the dynamic ranging frequency operation, which reduces the DL-TDOA ranging frequency during conditions of low NLOS probability, thus saving power. This decrease in ranging frequency does not, however, affect its localization accuracy, because of the augmented TDOA values supplied by the RNN-based localization predictor. Overall, we achieve $46.3\%$ lower power consumption compared to baseline methods. 

For the predominantly NLOS situation, the gap between the average current consumption of the proposed algorithm and Baseline 1 decreases to $2.6$~mA. 
This is because, in a predominantly NLOS situation, the dynamic ranging frequency algorithm does not reduce the ranging frequency as in the LOS situation. 
This is required to maintain sufficient localization accuracy in a NLOS scenario. 
Baseline 2 remains unaffected in terms of current consumption. Baseline 3, which uses a NLOS classifier-based kalman filter-adaptation, has to make repeated changes to the noise and covariance matrices of the filter, and thus has a higher current consumption in the NLOS case. 
The CNN-based NLOS probability prediction, the high ranging frequency, and the RNN-based localization prediction using augmented TDOA, all lead to the superior localization accuracy of the proposed algorithm. 
All three baseline algorithms are heavily affected by the NLOS-based localization inaccuracies. 
Baseline 3, with its deep learning-based NLOS classification and adaptive kalman filtering, is able to perform slightly better than baseline 2, which uses a combination of TDOA and TOF. Baseline 1, the least-squares-based estimation, is worst affected by the NLOS and multipath effects and has no interpolation or prediction capabilities to estimate future localization coordinates beforehand. Overall, we achieve $50.4\%$ lower localization error compared to baseline methods.

\section{Conclusions}
\label{sec:conclusion}
In this paper, we propose a novel channel-aware low-power UWB DL-TDOA localization algorithm. The proposed algorithm is scalable, because it does not involve one-to-one anchor-user device communications. It is highly accurate, as well as power-efficient by virtue of its dynamic ranging frequency setting. We obtain $50\%$ higher localization accuracy compared to baseline methods in predominantly NLOS conditions, and $46\%$ lower power consumption in LOS conditions.

\footnotesize
\bibliographystyle{IEEEtran}
\bibliography{main.bib}

\end{document}